\documentclass[prd,aps,twocolumn,showpacs,preprintnumbers,amsmath,amssymb]{revtex4}
\usepackage[dvips]{graphicx}
\usepackage{amsmath}
\usepackage{amssymb}

\usepackage{graphicx}
\usepackage{dcolumn}
\usepackage{bm}
\def\Dsl{\hbox{/\kern-.6700em\it D}} 
\def\dsl{\hbox{/\kern-.5300em$\partial$}}

\def\be{\begin{equation}}
\def\ee{\end{equation}}
\def\ba{\begin{eqnarray}}
\def\ea{\end{eqnarray}}
\def\nn{\nonumber}

\begin{document}

\preprint{HUTP-06/A0011}

\title{Inflation Free, Stringy Generation of Scale-Invariant Cosmological
Fluctuations in $D = 3 + 1$ Dimensions}
\author{Ali Nayeri}%
\email{nayeri@feynman.harvrad.edu}%
\affiliation{\qquad Jefferson Physical Laboratory, Harvard
University,Cambridge, MA 02138, USA}

\date{\today}

\pacs{98.80.Cq}

\begin{abstract}
We propose an alternative scenario to cosmic inflation for producing
the initial seeds of cosmic structures.  The cosmological
fluctuations are generated by thermal fluctuation of the energy
density of the ideal string gas in three {\it compact} spatial
dimensions. Statistical mechanics of the strings reveals that scalar
power spectrum of the cosmological fluctuations on cosmic scales is
scale-invariant for closed strings and inclines towards red for open
strings in three compact spatial dimensions. This generation of
thermal fluctuations happens during the Hagedorn era of string gas
cosmology and without invoking an inflationary epoch the
perturbations enter the radiation-dominated era.  The amplitude of
the fluctuations is proportional to the ratio of the two length
scales in the theory, i.e., the Planck length over the string
length, $(\ell_{Pl}/\ell_s)$.  Since modes with the shorter
wavelengths exit the Hubble radius at the end of the Hagedorn phase
at later times compare to the modes with long wavelengths, the
scalar fluctuations gain mild {\it tilt} towards {\it red}.
\end{abstract}

\maketitle

If the string theory is claimed to be the theory of every {\it
fundamental} things in the universe then it is better to describe
the cosmos itself. There have been many attempts to combine string
theory with cosmology.  One such approach is the string gas
cosmology (SGC) \cite{bv:1988,tv:1992}--also \cite{rbr1,rbr2} for an
overview of the subject, and \cite{str} for a critical review--which
is based specifically on new symmetries (T-duality) and new degrees
of freedom (string winding modes) of string theory \cite{bv:1988}
(see also \cite{Perlt}). Based on considerations of string
thermodynamics, it was argued that string theory could provide a
nonsingular cosmology. Going backwards in time, the universe
contracts and the temperature grows. However, the temperature will
not exceed the Hagedorn temperature for weakly interacting closed
strings. As the radius of space approaches the self-dual radius (the
string length), the pressure of the string gas will almost vanishes
because of T-duality (the positive contribution to the pressure from
momentum modes will cancel against the negative pressure from string
winding modes). Using the background equations of motion from
dilaton gravity, it follows that the evolution of the scale factor
near the self-dual radius will be quasi-static \cite{tv:1992}. Once
the radius of space decreases below the self-dual radius, the string
gas temperature will decrease, demonstrating that string gas
cosmology will be non-singular.  There has recently been quite a lot
of work on further developing string gas cosmology (see e.g.
\cite{ABE,Watson,Patil1,Patil2,Edna} and
\cite{RHBrev2,WatBatrev,RHBrev3} for recent reviews and
comprehensive lists of references).

In string gas cosmology, it is assumed that the universe starts in a
Hagedorn phase, a phase in which the universe is quasi-static and in
thermal equilibrium at a temperature close to the Hagedorn
temperature \cite{hagedorn:1965}, the limiting temperature of
perturbative string theory. As the universe slowly expands, heavy
degrees of freedom gradually fall out of equilibrium. String winding
modes keep all but $d$ spatial dimensions compact \cite{bv:1988}
(see, however, \cite{Columbia,DFM} for a critical view of this
aspect of the scenario). In this sense, one can view SGC as being a
stringy generalization of the big bang cosmology.

Hence SGC appears to offer a dynamical explanation for the
dimensionality of macroscopic spacetime in the context of a
non-singular cosmology. Independent of whether or not the mechanism
of \cite{bv:1988} is realized, SGC has more recently found a new
application as a possible solution of the moduli problem. In the
context of heterotic or bosonic string theory, it has been
demonstrated that massless string states which appear at the self
dual radius can stabilize all shape and radial moduli corresponding
to a toroidal compactification at the string scale
\cite{Watson,kanno}, in a way that is consistent with observational
bounds and late time cosmology \cite{Patil1,Patil2}.

On the other hand, one of the biggest triumph of the cosmic
inflation is explaining the origin of the structures by generating
quantum fluctuations in the very early universe
\cite{Guth:1980zm,Linde:1981mu,Linde:1982uu,Albrecht:1982wi,
Guth:1982ec,Linde:1983gd}. Despite many theoretical and
observational successes of the inflationary models, perhaps the most
unsatisfactory feature of all the inflationary models is how to
embed inflation in a more fundamental theory. For that, there have
been lots of attempts to find inflation in the context of string
theory in recent years \cite{Kachru:2003aw,Kachru:2003sx}.
Alternatively, one can ask whether a fundamental theory like string
theory can provide an alternative model of generating the initial
seeds for formation of structures
\cite{Steinhardt:2002ih,Nayeri:2005ck,Brandenberger:2006xi}.
Recently, a mechanism to generate an initial scale-invariant
spectrum of scalar metric fluctuations in the context of SGC  and
without invoking an inflationary period, has been proposed
\cite{Nayeri:2005ck}. It was demonstrated that, during the
quasi-static Hagedorn phase of SGC, thermal fluctuations of a closed
string gas generate an almost scale-invariant spectrum of metric
fluctuations for the gravitational potential, which is the quantity
which generates the observed anisotropies in the cosmic microwave
background (CMB). In this paper, I will elaborate more on the result
of \cite{Nayeri:2005ck} and compare the results with the ones for
open strings and massless relativistic particles.

In what follows, we shall assume that there are only $d$ already
sufficiently large spatial dimensions and will work in natural units
($c = \hbar = k_B = 1$)

\section{String Gas Cosmology with Dilaton: Dynamics}
String theory predicts the presence of a scalar field known as {\it
dilaton} which along with the graviton and the antisymmetric tensor
couples to matter.  The low energy string theory effective action in
the `string' frame is
\ba
\mathcal{A} & = & - \frac{1}{2\kappa_D^2}
\int \sqrt{-G} \,
d^{D}
x \, e^{- 2\phi} \left[^{(D)}R + 4 \nabla_\mu \phi \nabla^\mu \phi
\right. \nn \\
&&\left. - \frac{1}{12}H_{\mu\nu\lambda}H^{\mu\nu\lambda}
-\frac{1}{4} F_{\mu\nu}F^{\mu\nu}\right] +
\mathcal{A}_m\,,\label{effective_action}
\ea
where $^{(D)}R$ is the $D$-dimensional Ricci scalar, $\phi$ the
dilaton, $H_{\mu\nu\lambda} = \partial_{[\mu}B_{\nu\lambda]}$, the
field strength of the antisymmetric field tensor $B_{\mu\nu}$,
$F_{\mu\nu} = \partial_{[\mu} A_{\nu]}$, the field strength of the
$U(1)$ gauge field $A_\mu$ and ${\mathcal A}_m$ is the action for
string matter which will be discussed later.

If one rescales the metric in the following way
\be
G_{\mu\nu} \rightarrow e^{2[2\phi/(D - 2)]} g_{\mu\nu}\,,
\ee
the low energy effective action becomes a $D$ dimensional modified
Einstein action.  This transformation of the metric defines the
Einstein frame in which the corresponding action to
(\ref{effective_action}) takes the following form in this frame
\ba
\mathcal{A}_E & = & - \frac{1}{2\kappa_D^2} \int \sqrt{-g}\,
d^{D} x \left[^{(D)}R + \frac{2.2}{(D - 2)} \nabla_\mu \phi
\nabla^\mu
\phi \right. \nn \\
&&  - \frac{1}{12} e^{- 2[4\phi/(D - 2)]\phi}
H_{\mu\nu\lambda}H^{\mu\nu\lambda}  \nn \\
&& - \left.\frac{1}{4} e^{- 2[2\phi/(D - 2)]\phi}
F_{\mu\nu}F^{\mu\nu}\right] + \mathcal{A}_m\,.
\label{einstein_action}
\ea

If we now consider bosonic strings in $D = 10$ critical dimensions
by ignoring the antisymmetric tensor $B_{\mu\nu}$ for simplicity,
the action for the gravitational degrees of freedom interacting with
string matter then, in string frame, takes the form
\ba
 \mathcal{A} = && -
\frac{1}{2\kappa_{10}^2} \int \sqrt{- G}\,
d^{10} x \, e^{- 2\phi} \left[^{(10)}R + 4(\nabla\phi)^2 +...\right]
\nn
\\ && + {\mathcal A}_m\,,\label{dilaction}
\ea
where $\kappa_{10}^2 = \frac{1}{2} (2\pi)^7\ell_s^2$, and
\be
{\mathcal A}_m =  \int dt\; F(a, \beta)\,,
\ee
is the action for the {\it stringy} matter which consists of a gas
of almost ``free'' string modes, by assuming small effective string
coupling constant, in thermal equilibrium at the temperature
$\beta^{-1}$.  Here $F$ is the (one-loop) free energy which can be
expressed in terms of the one-loop string partition function in a
torus background of radii $a$ and periodic Euclidean time of
perimeter $\beta$.

For the spatial rectangular torus background with two isotropic sets
of dimensions of the form
\be
ds^2 = - d t^2 + \ell_s^2\left[\sum_\imath^{(d)} a^2(t)
d\theta_\imath^2 + \sum_\imath^{(9 - d)} a_s^2(t)
{d\theta'}_\imath^2\right] \,,
\ee
following equations of motion are found \cite{tv:1992}
\be
-(d)\dot \mu^2 - (9 - d) \dot\nu^2 + \dot\varphi^2 = 2
\frac{\kappa_{10}^2}{(2\pi \sqrt{\alpha'})^9} e^\varphi E\,,
\label{rescaledil_1}
\ee
\be
\ddot \mu - \dot\varphi \dot\mu  = \frac{\kappa_{10}^2}{(2\pi
\sqrt{\alpha'})^9} e^\varphi P_d\,, \label{rescaledil_2}
\ee
\be
\ddot\nu - \dot\varphi \dot\nu = \frac{\kappa_{10}^2}{(2\pi
\sqrt{\alpha'})^9} e^\varphi P_{9-d} \,, \label{rescaledil_3}
\ee
\be
\ddot\varphi - (d) \dot\mu^2 -(9 - d) \dot\nu^2 =
\frac{\kappa_{10}^2}{(2\pi \sqrt{\alpha'})^9} e^\varphi E
\label{rescaledil_4} \,,
\ee
or alternatively, in terms of $\phi$
\be
-(d)\dot \mu^2 - (9 - d) \dot\nu^2 + [\dot\phi - (d) \dot\mu -
(9 - d) \dot\nu ]^2 = 2 \kappa_{10}^2 e^{2\phi} \rho
\label{dil_1}\,,
\ee
\be
\ddot \mu - [\dot\phi - (d) \dot\mu - (9 - d) \dot\nu ] \dot\mu
= \kappa_{10}^2 e^{2\phi} p_d\,, \label{dil_2}
\ee
\be
\ddot\nu -  [\dot\phi - (d) \dot\mu - (9 - d) \dot\nu ]\dot\nu =
\kappa_{10}^2 e^{2\phi} p_{9-d} \,, \label{dil_3}
\ee
\be
[\ddot\phi - (d) \ddot\mu - (9 - d) \ddot\nu ] - (d) \dot\mu^2
-(9 - d) \dot\nu^2 = \kappa_{10}^2 e^{2\phi} \rho \,. \label{dil_4}
\ee
Here $^{(10)}R$ is the ten-dimensional Ricci scalar, $\phi$ is the
10-dimensional dilaton of the string theory which is related to the
shifted dilaton through
\be
\sqrt{G} e^{- 2 \phi} = e^{- \varphi} \,,
\ee
or
\be
\varphi \equiv 2 \phi - (d) \mu - (9 - d) \nu \,,
\ee
where
\be
a(t) = e^{\mu(t)} \,,\,\,\,\,\,\,\,\,\,\,\,\,\, a_s(t) =
e^{\nu(t)}\,,
\ee
are the homogenous scalar factors for $d$ expanding dimensions and
$9 - d$ dimensions remaining in string scale, respectively.  $\nu$
is normally taken to be zero.

The total energy of the system is represented by $E \equiv F + \beta
(\partial F/\partial \beta)$. The variables $P_d \equiv - (\partial
F/\partial \mu)$ and $P_{9 - d} \equiv - (\partial F/\partial \nu)$
are related to the pressures $p_d$ , $p_{9 - d}$ in the respective
directions by a volume rescaling, $P_X = p_X V$, with $V$ being the
total volume of the system given by
\be
 V = (2\pi\sqrt{\alpha'})^9 a^{(d)} a_s^{(9-d)} \equiv
(2\pi\sqrt{\alpha'})^9 e^{(d)\mu} e^{(9 - d)\nu} \,.
\ee
Therefore we can find the modified conservation law of
energy-momentum
\be
\dot E + (d) \dot\mu P_d+ (9 - d)\dot\nu P_{9 - d} = 0 =
\frac{\dot S}{\beta} \,,
\ee
or
\be
\dot \rho + (d)\frac{\dot a}{a}(\rho + p_d) + (9 - d)\frac{\dot
a_s}{a_s}(\rho + p_{9 - d}) = 0  \,,
\ee
where $S \equiv \beta^2 (\partial F/\partial \beta)$, the total
entropy, is conserved due to the fact that $F =
F[\mu(t),\nu(t),\beta(t)]$. The details of this free energy will be
discussed later.  The adiabaticity condition tells us that the
temperature $\beta^{-1}$ adjusts itself such that the total entropy,
$S$, remains constant for a given radii determined by $\mu$ and
$\nu$.

The above equations of motion are duality invariant since $F$ is
invariant under $\mu \rightarrow -\mu$ , $\nu \rightarrow -\nu$ and
$\varphi \rightarrow \varphi$ (or $\phi \rightarrow \phi - \mu -
\nu$) for a given temperature. Note that while $E(\mu,\nu)$ and
$\beta(\mu,\nu)$ are invariant under the duality transformations,
the pressures $P_d$ and $P_{9-d}$ change sign under these
transformations.

The system of equations (\ref{rescaledil_1})-(\ref{rescaledil_4}) or
(\ref{dil_1})-(\ref{dil_4}) has a mechanical interpretation of
describing a motion of a particle either in the potential
$U_\varphi(\mu,\nu,\varphi) = e^\varphi E$ or $U_\phi(\mu,\nu,\phi)
= e^\phi \rho$. Based on the behavior of $\varphi$ ($\phi$) and $E$
($\rho)$ one can distinguished two distinct regime for the system of
a very weekly coupled string gas.  I will assume hereafter the
expansion/contraction only exists in $d$ dimensions and that the $(9
- d)$ dimensions have stretched (contracted) to their maximum
(minimum) length, i.e., string (Planck) length and thus they do not
take part in expansion (contraction).  Hence, $\nu = 0$ or $\dot a_s
= 0$.

{\bf Hagedorn phase.} For weakly coupled strings, $g_s \ll 1$, there
is a limiting temperature known as the Hagedorn temperature
\cite{hagedorn:1965}.  In this regime, because of weak interaction
among the strings, to a good approximation the free energy of the
string gas, $F = E - T_H {\mathcal S}$, vanishes (since for the
string gas ${\mathcal S} \approx \beta_H E$) and so does the total
pressure $P = P_d = - (\partial F/\partial \ln{a}) \approx 0$.  The
equations of motion (\ref{rescaledil_1}), (\ref{rescaledil_4}) and
(\ref{rescaledil_2}) for this vacuum reduce to
\be
\ddot \varphi = \frac{1}{2}\left[{\dot\varphi}^2 +
(d){\dot\mu}^2 \right]\,,
\ee
\be
\ddot \mu - \dot\varphi \dot\mu = 0 \,.
\ee
Of course, one trivial solution to the above equations is for
$\varphi$ and $\mu$ to be constant but since the total energy, $E$,
in the absence of the total pressure, is conserved, one can also
find non-trivial solutions to the above equations:
\be
\varphi(t) = \varphi_0 +
\ln{\left[\frac{2(2\pi\sqrt{\alpha'})^9}{(\kappa_{10}^2 E)}
\frac{1}{t(t - t_0)}\right]}\,,
\ee
\be \mu(t) = \mu_0 + \mu_1 \ln{\left[\frac{t}{t - t_0}\right]} \,.
\ee

\begin{figure}
\includegraphics[height=5cm]{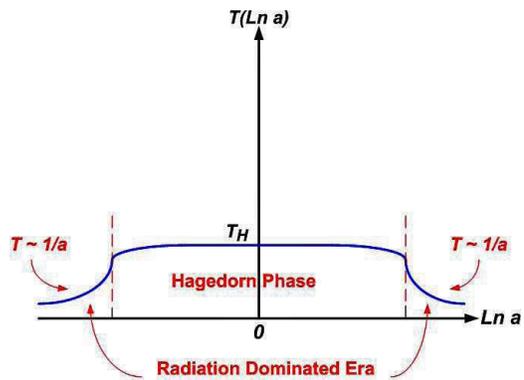}
\caption{Temperature is invariant under transformation of $a
\rightarrow (1/a)$.  The closer we get to the Hagedorn temperature,
which depends on the total entropy of the universe, the Hagedorn
phase would be longer leaving enough time for the strings to
homogenize the universe and monopoles annihilation.} \label{fig:1}
\end{figure}

Note that, since $E$ is positive (ignoring Casimir-type
contributions), from (\ref{dil_1}) one finds that $\dot\varphi$
never changes sign, i.e., never becomes zero. Assuming forward time,
one can choose damping effect for $\dot\varphi$ which corresponds
choosing $\dot\varphi < 0$ to ensure the adiabacity approximation in
driving the equations of motion.  Combing this with $\dot\mu > 0$
guarantees that we are in the weak-coupling regime as we have
assumed.  Also Eq. (\ref{dil_1}) imposes the condition $t_0^2(1 -
3\mu_1^2) = 0$ which has two implications either $t_0 = 0$ and thus
the universe is static (since the scale factor would be time
independent) or $\mu_1^2 = (1/3)$ and the universe expands.  Either
way, at late times, the dilaton asymptotes a constant value and
$$
a(t) \rightarrow \mbox{constant} \Rightarrow H \equiv \frac{\dot
a}{a} \rightarrow 0\,.
$$
{\bf Radiation dominated era.}  As the universe slowly expands,
$\mu(t)$ reaches the intermediate region in which $E$ drops and the
system moves towards the direction of large $\mu$.  In this regime
the dilaton $\phi$ is a constant and thus one can easily find out
that
\be
P = \left(\frac{1}{d}\right) E\,,
\ee
with $E = E_0 e^{-\mu}$ and thus
\be \rho = \rho_0 e^{-(d + 1)\mu}\,. \ee
This is a description of a matter with traceless energy-momentum
tensor, i.e., a gas of massless particles in a thermodynamical
equilibrium.   In this regime, a power law solution for $\mu$ and
$\varphi$ exists in the form of
\be
 \mu(t) = \mu_0 + \left(\frac{2}{d + 1}\right)\ln{t} \,,
 \ee
\be
\varphi(t) = \varphi_0 - \left(\frac{2 d}{d + 1}\right)\ln{t}\,,
\ee
which implies that
\be
\phi = \frac{1}{2}[\varphi + (d) \mu] = \mbox{constant} \,,
\ee
and
$$
a(t) \sim t^{2/(d + 1)} \rightarrow H(t)\sim \frac{2}{d + 1}
\frac{1}{t} \,.
$$

To summarize, the universe, in the expanding sector, starts out in a
quasi-static phase known as the Hagedorn phase in which the
thermodynamical equilibrium can be established over the entire
universe.  In this regime the universe is large enough so that after
shrinking to a microscopic scale ($H^{-1} \sim
\ell_s^2/\ell_{Pl}^2$), in the follow up radiation dominated era,
can grow to the present size without undergoing through any
inflationary phase. The phase transition to the radiation dominated
FRW universe occurs when the dilaton gets fixed at the end of the
Hagedorn phase.

We want to study the statistical mechanics of the string gas in the
Hagedorn regime in which the coordinate radii are all near the
string scale, $\sqrt{\alpha'}= \ell_s$.  I first give a quick review
of the fundamentals of statistical mechanics that we need later.

\section{Brief Review of Fundamental of Statistical Mechanics}

For any thermodynamical system there are two almost identical
statistical descriptions.  The {\it microcanonical ensemble} is the
most basic description of statistical mechanics and is valid for a
closed system with fixed total energy $E$ (and volume $V$).  The
fundamental quantity in the ensemble is the total density of states
$\Omega(E)$ which is defined as
\be
\Omega(E) \equiv \sum_{\imath} \delta (E - E_\imath)\,,
\ee
where the sum is over all states $\imath$ of the system and
$E_\imath$ denotes the energy of the state $\imath$.  The quantity
$\Omega(E)$ is related to the thermodynamic entropy $S$ of the
system, up to an additive constant, as
\be
S(E) \equiv \ln{\Omega(E)}\,.
\ee
Once the entropy is determined, other thermodynamical quantities
like temperature, $T$, pressure, $P$ and the specific heat $C_V$ can
be defined through the first law of thermodynamics $dE = T dS - P
dV$. Thus
\be
\frac{1}{T(E)} = \beta (E)  \equiv \left(\frac{\partial
S}{\partial E}\right)_V = \left(\frac{\partial \ln{\Omega}}{\partial
E}\right)_V \,,
\ee
\be
P \equiv T \left(\frac{\partial S}{\partial V}\right)_E = T
\left(\frac{\partial \ln{\Omega}}{\partial V}\right)_E \,,
\ee
and
\be
 C_V  \equiv  \left(\frac{\partial E}{\partial T}\right)_V =
-\beta^2 \left(\frac{\partial E}{\partial \beta}\right)_V  = -
\left[T^2\left(\frac{\partial^2 S}{\partial
E^2}\right)_V\right]^{-1}\,. \ee
On the other hand, the {\it canonical ensemble} can be used to
describe the properties of any large subsystem of a closed system,
providing the energy is an extensive variable.  The basic quantity
in this ensemble is the canonical partition function
$$
Z(\beta) = \sum_\imath e^{-\beta E_\imath} \,,
$$
which is the Laplace transformation of the total energy density
$\Omega(E)$ for any system whose energy is bounded from below (say,
unlike the gravitating system),
\be
Z(\beta) = \int_0^\infty d E e^{-\beta E}
\Omega(E)\,,\label{laplace}
\ee
or inversely, if $Z(\beta)$ exists and is known as a function of
$\beta$ then $\Omega(E)$ can be obtained as a function of $E$ from
the inverse Laplace transformation,
\be
\Omega(E) = \int_{L - i\infty}^{L + i \infty} \frac{d\beta}{2\pi
i} e^{\beta E} Z(\beta) \,, \label{inverse_laplace}
\ee
where $L (= \mathfrak{Re}\beta)$ is the contour which is chosen to
be right of all the singularities of $Z(\beta)$ in the complex
$\beta$ plane.

Note that while (\ref{inverse_laplace}) always exists, the
transformation (\ref{laplace}) is subject to the existence of the
saddle point.

Both canonical and microcanonical descriptions of statistical
mechanics can be used in an interchangeable manner for describing
the gaseous systems.

While for a system that is described by microcanonical distribution,
the total energy $E$ is fixed, the canonical distribution describes
the same system in a way that the total energy is not a fixed
quantity but rather fluctuates among the canonical ensemble.  If we
should be able to use either of the two descriptions, in an
interchangeable manner, then to describe the system one must
satisfies the following two conditions: (a) The mean energy $\langle
E \rangle$ of the canonical ensemble must be the same as the fixed
energy of the microcanonical ensemble and (b) the root-mean-square
fluctuations in the energy, in the canonical distribution, $\Delta
E/\langle E \rangle \equiv \sqrt{\langle E^2 \rangle /\langle E
\rangle^2 - 1}$, must be negligible for sufficiently large systems,
i.e., $\Delta E/\langle E \rangle \propto 1/\sqrt{\langle E
\rangle}$. Since in the canonical distribution, $\langle E \rangle
\equiv - Z^{-1}(\partial Z/\partial \beta) = -(Z^\prime/Z)$ and
$\langle E^2 \rangle \equiv Z^{-1}(\partial^2 Z/\partial \beta^2) =
(Z''/Z)$, where the primes denote the differentiation with respect
to $\beta$, it follows that
\ba C_V & = & \beta^2 \left(\langle E^2 \rangle - \langle E
\rangle^2 \right) = \beta^2\left(\frac{Z''}{Z} -
\frac{Z'^2}{Z^2}\right) \nn \\
& = &\beta^2 \frac{\partial}{\partial
\beta}\left(\frac{Z'}{Z}\right)= - \beta^2 \frac{\partial \langle E
\rangle}{\partial \beta} \,. \ea
The specific heat defined in the canonical distribution must,
therefore, be positive definite.  As long as the two distributions
are equivalent the specific heat in microcanonical description will
be positive definite.  There are regimes in which the specific heat
in microcanonical distribution is negative and thus the equivalence
between the two prescriptions breaks down.  One example of this
situation is near the phase transition, where the fluctuations may
become very large.

\section{Density of States for the Ideal String Gas}
When the string coupling is sufficiently small, $g_s \ll 1$, and the
local spacetime geometry is close to flat $\mathbb{R}^{d + 1}$ over
the length scale of the finite size box of volume $V = R^d$, there
are two distinct regimes that characterizes the statistical
mechanics of string thermodynamics: the massless modes with field
theoretic entropy, $S \propto E^{d/(d + 1)}$, and the highly excited
strings with $S \propto E$. One simple realization of this setup for
a string background is a spatial toroidal compactification with $d$
dimensions of size $R$ and $9 - d$ dimensions of string scale size.
Small string coupling ensures us that we can measure energies with
respect to the flat time coordinate.

To review the features of string thermodynamics we use the intuitive
geometrical picture for a highly excited string as a random walk in
target space \cite{abkr:1999}.  The large entropy factor
corresponding to the shape of the random walk in space explains why
highly energetic strings dominate the thermodynamics despite of
their large energy.

\textbf{Closed strings.} For a highly excited closed string
represented as a random walk in a target space, the energy
$\varepsilon$ of the string is proportional to the length of the
random walk.  Thus the number of the strings with a fixed starting
point grows as $\exp{(\beta_H \varepsilon)}$, with $T_H = 1/\beta_H$
being the Hagedorn temperature. This explains the bulk of the
entropy of highly energetic strings. Closed strings correspond to
random walks that must close on themselves.  This overcounts by a
factor of roughly the volume of the walk, denoted $V_{\rm
walk}(\varepsilon)$. The global translation of the random walk in
volume $V = R^d$ and $1/\varepsilon$ due the fact that any point in
the string can be considered as a new starting point are other
factors that contribute to the number of closed string. Therefore,
the final result is
\be
\omega_{\textrm{closed}}(\varepsilon) \sim V .
\frac{1}{\varepsilon} . \frac{e^{\beta_H \varepsilon}}{V_{\rm
walk}(\varepsilon)}\,.
\ee
There are two limiting case here: \\
\textbf{(a)} Volume of the random walk is  well-contained in $d$
spatial dimensions (i.e., $R \gg \sqrt{\varepsilon}$) which
corresponds to a string in $d$ non-compact dimensions.  In this case
the $V_{\rm walk}(\varepsilon) \sim \varepsilon^{d/2}$, the density
of states per unit volume is
\be
\omega_{\rm{closed}}(\varepsilon) \sim R^d \frac{e^{\beta_H
\varepsilon}}{\varepsilon^{d/2 + 1}} \,.
\ee
\textbf{(b)} Volume of the random walk is space-filling ($R \ll
\sqrt{\varepsilon}$) and saturates at order $V$ which corresponds to
$d$ compact dimensions that contains the highly excited string
states. Hence,
\be
\omega_{\rm{closed}}(\varepsilon) =  \frac{e^{\beta_H
\varepsilon}}{ \varepsilon}\,.
\ee
This is an exact leading result.  Note that here the density of
states are almost independent of the topology of the spacetime and
depends only on the volume of the random walk.  One can combine the
two regimes in one equation
\be
\omega_{\rm{closed}}(\varepsilon) = \beta_H R^d \frac{e^{\beta_H
\varepsilon}}{(\beta_H\varepsilon)^{\gamma_c + 1}}\,, \mbox{with
$\gamma_c = \frac{d_c}{2}$\,,}
\ee
where $d_c$ is the number of dimensions in which closed strings have
no windings and again $R^d$ is the volume of this space.  In other
words, at low energies the winding modes are frozen and thus
corresponds to large radius, while at high energies the the winding
modes can be excited and thus all the radii are compact.  We want to
find the total density of states when the radius of the compact
space expands and thus one can see the effect of interplay between
the winding modes and the momentum modes of the closed strings.

The total density of states, $\Omega (E) $, can be obtained through
(\ref{inverse_laplace}).  The partition function $Z(\beta)$ can be
evaluated explicitly in the one loop approximation \cite{joe:1986}
\be
Z(\beta, R) = \sum_\alpha e^{-\beta E_\alpha(R)} \,,
\ee
where $\alpha = (N,q_1, \dots, q_N)$ labels a state for $N$ strings
and $q_k$ are the quantum numbers of the $k^{th}$ string and stand
for the whole set of momentum, winding and oscillatory modes.  Here
$E_{\alpha}$ is the energy of the multi-string state $\alpha$ in a
universe of radius $R$.  In the ideal gas approximation, $E_\alpha$
is given by the sum all over the single-state energies, i.e.,
$E_\alpha(R) = \sum_{k = 1}^N \varepsilon_{q_k}(R)$, where
$\varepsilon$, say, for the closed bosonic strings is
\cite{gsb:1982},
\be
\varepsilon^2 = \frac{\mathbf{l}^2}{R^2} + \frac{\mathbf{w}^2
R^2}{\ell_s^4} + \frac{2}{\ell_s^2}\left[-2 + \sum_{I = 1}^{d
-1}\sum_{m = 1}^\infty m(N_m^I +
\tilde{N}_m^I)\right]\,,\label{e_spectrum}
\ee
which maps to itself under the duality transformations
$$
R \longleftrightarrow \tilde{R} = (\ell_s^2/R) \,,\,\,\,\,\mathbf{l}
\longleftrightarrow \mathbf{w}\,.
$$
In (\ref{e_spectrum}), $\mathbf{l}$, $\mathbf{w}$, $N_m^I$ and
$\tilde{N}_m^I$ are momentum, winding and oscillatory quantum
numbers, respectively.

Near the Hagedorn temperature,  we can assume Maxwell-Boltzman
statistics and thus treat the system quasiclassically.  Then we can
write $Z(\beta) = \exp{[z(\beta)]}$, where $z(\beta)$ is the
single-string partition function (free thermal energy)
\be
z(\beta) = \int_0^\infty d\varepsilon \omega(\varepsilon)
e^{-\beta \varepsilon} = \sum_q e^{-\beta \varepsilon_q}\,.
\ee
According to \cite{djt:1989a,djt:1989b,djnt:1992}, the singular part
of the partition function at finite volume for closed strings is
given by a set of poles of even multiplicity $g_\imath = 2 k_\imath
= 2 d$
\be
Z^{\rm singular}_{\imath,closed} \sim
\left(\frac{\beta_\imath}{\beta - \beta_\imath}\right)^{k_\imath}
\,, \label{Z_singgular}
\ee
with $k_\imath = k_0 = 1$ for the leading Hagedorn singularity
$\beta_\imath = \beta_0 =  \beta_H = \sqrt{2} \pi \ell_s
(\sqrt{\omega_l} + \sqrt{\omega_r})$ where $(\omega_l,\omega_r)$ is
$(2,1)$,$(2,2)$ and $(1,1)$ respectively for bosonic, closed and
heterotic strings. Other singularities are located to the left of
$\beta_H$ at
\be
\beta_\imath = \sqrt{2} \pi \ell_s \left[\left(\omega_l -
\frac{\imath}{2}\frac{\ell_s^2}{R^2}\right)^{\frac{1}{2}} +
\left(\omega_r -
\frac{\imath}{2}\frac{\ell_s^2}{R^2}\right)^{\frac{1}{2}}\right]\,,
\label{beta_expansion}
\ee
for $\imath = 1, 2, \dots, \mathcal{O}(R/\ell_s)$ or
\be
\beta_\imath = \sqrt{2} \pi \ell_s \left[\left(\omega_l -
\frac{\imath}{2}\frac{R^2}{\ell_s^2}\right)^{\frac{1}{2}} +
\left(\omega_r -
\frac{\imath}{2}\frac{R^2}{\ell_s^2}\right)^{\frac{1}{2}}\right]\,,
\label{beta_collaps}
\ee
for $\imath = 1, 2, \dots, \mathcal{O}(\ell_s/R)$, depending on
which one has a larger real part.  In other words, while
(\ref{beta_expansion}) shows the location of the singularities for
an expanding universe, the singularities of its dual contracting
universe are given by (\ref{beta_collaps}) and vise vera.

The regular part of the free energy to the leading order in energy
is
\be
Z^{\rm regular}_\imath \sim n_H V - \rho_H V (\beta -
\beta_\imath) + {\cal O}(V(\beta - \beta_\imath)^2) \,.
\label{z_reg}
\ee
The leading singularity at very high and finite volume is always a
simple pole of the partition function at the Hagedorn singularity.
Considering the subleading singularities that are located to the
left of $\beta_0 = \beta_H$, e.g., $\beta_1$, one can parameterize
$Z(\beta)$ in the region in which there are two singularities,
\be
Z_{closed}(\beta) \simeq   \frac{\beta_H}{(\beta -
\beta_H)}\left(\frac{\beta_H - \beta_1}{\beta - \beta_1}\right)^{2
d} . Z^{\rm regular}(\beta) \,.
\ee
The total density of states of closed strings when all the
dimensions are compact and large is \cite{djnt:1992}
\ba
\Omega(E , R)& \simeq & \Omega_0 + \Omega_1 \nn \\
& \simeq & \beta_H e^{\beta_H E + n_H V}[1 + \delta \Omega_{(1)}(E ,
R)] \label{density_states}\,, \ea
with $\Omega_0$ and $\Omega_{(1)}$ being, respectively, the
contributions to the density of states from $\beta_0 = \beta_H =
(1/T_H)$ and the closest singularity to $\beta_H$, i.e., $\beta_1 <
\beta_H$ (which is real).  The subleading contributions are encoded
in  $\delta \Omega_{(1)}$ which is
\be
 \delta \Omega_{(1)}(E , R) = - \frac{(\beta_H E)^{2d - 1}}{(2d -
1)!} e^{-(\beta_H - \beta_1)(E  - \rho_H V)} \,. \ee

Here, the density of states has specifically written for $d$ large
compact dimensions, i.e., $R \gg \ell_s$.  Also, $n_H$ is a constant
number density of order $\ell_s^{-d}$ and $\rho_H$ is the `Hagedorn
Energy density' of the order $\ell_s^{-(d + 1)}$ while according to
(\ref{beta_expansion}) and (\ref{beta_collaps})
\be
\beta_H - \beta_1 \sim \left\{ \begin{array}{ll}
(\ell_s^3/R^2) \,, & \mbox{for $R \gg \ell_s$}\,, \nonumber \\
(R^2/\ell_s)\,, & \mbox{for $R \ll \ell_s$}\,.
\end{array}
\right.
\ee
To ensure the validity of Eq. (\ref{density_states}) we demand that
$- \delta \Omega_{(1)} \ll 1$ by assuming $\rho \equiv (E / V) \gg
\rho_H$. So the entropy of the string gas in the Hagedorn phase is
given by
\be
S(E , R) \simeq \beta_H E + n_H V + \ln{\left[1 + \delta
\Omega_{(1)}\right]} \,,
\ee
and therefore the temperature $T \equiv [(\partial S/\partial
E)_V]^{-1}$ will be
\ba
T(E , R) & \simeq & \left(\beta_H + \frac{\partial \delta
\Omega_{(1)}/\partial E}{1 + \delta \Omega_{(1)}}\right)^{-1}  \nonumber \\
& \simeq & T_H \left(1 + \frac{\beta_H - \beta_1}{\beta_H} \delta
\Omega_{(1)}\right)\label{temp}\,.
\ea

\textbf{Open Strings.}  For the open strings, however, the
geometrical picture is a bit more involved.  Let's consider a highly
excited string between $Dp$- and $Dq$-branes.  In addition to the
leading exponential degeneracy for random walks with a fix starting
point on the $Dp$-brane, there would be a degeneracy factor due to
the fixing of the endpoints on each brane
$$
(V^{\rm walk}_{NN}V^{\rm walk}_{ND}).(V^{\rm walk}_{NN}V^{\rm
walk}_{DN})\,,
$$
where $N$ and $D$ refer to Neumann and Dirichlet boundary
conditions.  Finally there would be an overall factor due to the
translation of the walk in the excluded $NN$ volume which is
$(V_{NN}/V^{\rm walk}_{NN})$.  Thus for the open string, the density
of the states looks like
\ba \omega_{\rm{open}}(\varepsilon) & \sim & \frac{V_{NN}}{V^{\rm
walk}_{NN}}.(V^{\rm walk}_{NN}V^{\rm walk}_{ND}).(V^{\rm
walk}_{NN}V^{\rm walk}_{DN}).\frac{e^{\beta_H \varepsilon}}{V^{\rm
walk}_{open}}\nonumber \\
&\sim & \frac{V_{NN}}{V^{\rm walk}_{DD}} e^{\beta_H \varepsilon} \,,
\ea
where $V^{\rm walk}_{\rm{open}} = V^{\rm walk}_{NN}.V^{\rm
walk}_{ND}.V^{\rm walk}_{DN}.V^{\rm walk}_{DD}$ is the total volume
of the random walk.  Like the close string there are two limiting case here
\\
\textbf{(a)}  The random walk is well-contained in the $d_{DD} =
d_\perp$ directions with $DD$ boundary conditions ($R_{DD} = R_\perp
\gg \sqrt{\varepsilon}$) we have $V^{\rm walk}_{DD} \sim \varepsilon
^{d_\perp/2}$ and thus
\be
\frac{\omega_{\rm{open}}}{V_{NN}} \sim \frac{e^{\beta_H
\varepsilon}}{\varepsilon^{d_\perp/2}}\,.
\ee
\textbf{(b)} If the random walk is filling the $DD$ volume then
\be \frac{\omega_{\rm{open}}}{V_{NN}} \sim \frac{e^{\beta_H
\varepsilon}}{V_{DD}}\,. \ee
In summary the density of states for a single open string can be
written as \cite{abkr:1999,lt:1995,lt:1997}
\be \omega_{\rm{open}}(\varepsilon) =  \beta_H V_o
\frac{V_{\|}}{V_\perp} \frac{e^{\beta_H \varepsilon}}{(\beta_H
\varepsilon)^{\gamma_o + 1}} \,, \mbox{with $\gamma_o =
\frac{d_o}{2} - 1$\,,} \ee
where $V_\| = V_{NN}$ and $V_\perp = V_{DD}$ are the volumes
transverse and perpendicular to the $D$-brane.  $0 \leq d_o \leq
d_\perp = d_{DD}$ is the number of dimensions transverse to the
brane with no windings and $V_o$ is the volume of this space (which
is $\mathcal{O}(1)$ in string units when there are windings in all
directions) .  Both $\gamma_o$ and $\gamma_c$ are
$\varepsilon$-dependent critical exponent.  The `effective' number
of large spacetime dimensions (i.e, the total number of $NN + DD$
dimensions) as a function of $\varepsilon$ is,
\be d_{o_{eff}}(\varepsilon) = d_{NN} + d_o(\varepsilon) \,, \ee
with $d_{NN} = p + 1$ being the $p$ spatial non-compact Neumann
directions of the $Dp$-brane.  In particular, if we consider 3 + 1
directions which are much larger than the string scale, then there
are $d_{comp}$ compactified dimensions of the string scale and
$d_{DD} = 6 - d_{comp}$ compactified directions much smaller than
the string scale.  For the large internal energies open strings can
move freely in the entire space  and we have $\gamma_o + 1 = 0$. In
the low energies limit, on the other hand, $\gamma_o + 1 = d_{DD}/2
= 3 -  d_{comp}/2$.

The behavior of $z^{\rm singular}_{\imath, open}(\beta)$ for open
strings near the singularity $\beta \approx \beta_\imath$, by direct
substitution, is given by \cite{abkr:1999}
\begin{widetext}
\be
z_{\imath, open}^{\rm singular}(\beta) \sim\left\{
\begin{array}{ll}
\frac{1}{2} \frac{(-1)^{\gamma_o + 1}}{\Gamma(\gamma_0  + 1)} f
(\beta - \beta_\imath)^{\gamma_o} \log{(\beta - \beta_\imath)} \,, &
\mbox{for $\gamma_o \in {\cal Z}^+ \cup \{0\}$}\,,  \\
\frac{1}{2} \Gamma(-\gamma_o) f (\beta - \beta_\imath)^{\gamma_o}
\,, & \mbox{for $\gamma_o \in {\hskip -0.33cm \slash} \hskip +0.15cm
{\cal Z}^+ \cup \{0\}$}\,,  \label{z_sing}
\end{array}
\right.
\ee
\end{widetext}
where $f = V_\|/V_\perp = V_{NN}/V_{DD}$ is the volume factor.  The
critical exponent for the compact $DD$ directions $\gamma_o = -1$.
If a number $d_\infty$ of $DD$ directions are strictly non-compact,
then $\gamma_o \rightarrow \gamma_o + d_\infty/2$ and $f \rightarrow
f.V_\infty$.

Eq. (\ref{z_reg}) is a generic result for both closed and open
strings.  For open strings $V_o = V_\|$.

Whenever the specific heat is positive (and large), there is a
correspondence between the canonical and microcanonical ensembles
and thus the saddle point approximation is applicable.  A necessary
condition for this is that $\gamma < 1$, ensuring the canonical
internal energy $E(\beta) \sim \partial_\beta z(\beta)$ diverges at
the Hagedorn singularity. In other words, these systems are unable
to reach the Hagedorn temperature since their require an infinite
amount of energy to do so.  For these systems the Hagedorn
temperature is limiting, and this is true for all the open strings
with $d_{NN} > d_{o_{eff}} - 4$. For the closed strings the Hagedorn
temperature is non-limiting for any model in which $d_c
> 3$.  In other words, stable canonical (i.e., no phase transition)
can be achieved for the closed strings in low dimensional
thermodynamical limits $d_c \leq 2$, or open strings with $d_\perp
\leq 4$ non-compact $DD$ dimensions, i.e., $Dp$-brane with $p \geq
5$ and non-compact transverse dimensions.

Systems with close-packing of random walks (high energy in a fixed
volume) have $\gamma_c = 0$ for closes strings and $\gamma_o = -1$
for open strings.  As it was mentioned above in these cases we can
use the saddle point approximation.  For a gas of open strings we
have~\cite{abkr:1999}
\begin{widetext}
\be \Omega_{open}(\gamma_o = -1)  \simeq \beta_H f x^{-1} I_1(2x)
e^{\beta_H E + a_H V_\|}\left[1 + {\cal O}\left(\frac{x^2}{V_\|(\rho
- \rho_H)^2}\right) + {\cal O}\left(e^{-(\beta_H - \beta_1)(E -
\rho_H V_\|)}\right)\right] \,, \ee
\end{widetext}
where $x = \sqrt{f(E - \rho_c V_\|)}$ is the control parameter for
saddle point approximation and $I_1$ is the modified Bessel function
of the first kind. For $x \gg 1$ the Hagedorn temperature is
limiting and thus:
\be
\Omega_{open}(E) \sim \exp{\left(\beta_H E + 2\sqrt{f E}\right)}
\label{total_open} \,,
\ee
for $\gamma_o = -1$ and hence $d_{o_{eff}} = d_{NN}$.

The validity of the results in Eqs. (\ref{total_open}) and
(\ref{density_states}) depends on the condition
\be
\log{\left(\frac{\Omega_0}{\Omega_1}\right)} \gg 1 \,,
\ee
where $\Omega_1$ is the contribution to the density of states from
the closest singularity to $\beta_H$, i.e., $\beta_1 < \beta_H$
which is real.  Thus the necessary condition is
\be
\beta_H E + a_H V_\| \gg \beta_1 E + {\rm Re}(a_1) V_\|
\ee
or,
\be
(\beta_H - \beta_1)(E - \rho_H V_\|) \gg 1 \Rightarrow (\rho -
\rho_H) R^{D - 3} \gg 1 \,,
\ee
which is satisfied for large $R$ and high $E$, provided $D > 3$.

\section{Power Spectrum of the Energy Fluctuations}
Now, having the total density of states, we are in the position to
evaluate the power spectrum of our ideal gas in this 9-torus.

The entropy of the open and closed string system in the Hagedorn
temperature limiting case for large $R$ and $\beta_H E \gg 1$ is
\be
S_{\textrm{closed}}(E)  \approx  \beta_H E + n_H V_{D - 1} +
\ln{[1 + \delta \Omega_{(1)}]} \,, \label{s_closed}
\ee
and
\be
S_{\textrm{open}}(E) \approx  \beta_H E + n_H V_\| + 2
\sqrt{f\,E} \label{s_open} \,.
\ee
This clearly shows that the asymptotic entropy of the system of open
strings dominates over the closed strings in the finite volume
(compact dimensions):
$$
S_{\textrm{open}}(E) \gg S_{\textrm{closed}}(E) \,.
$$
It is worth mentioning that the entropy for an ideal gas of
relativistic point particles with energy density $\rho$ in a large
volume $V$ in $d$ spatial dimensions is
\be
S_{\textrm{particle}} \approx V \rho^{d/(d + 1)} \,.
\ee
Note that unlike the string case the coefficient of $V$ is not just
a constant, $n_H$, but the energy density, $\rho$.

By using the microcanonical density of states we can find the
temperature ($\beta = \partial_E S(E)$),
\be
\frac{1}{T} \approx \frac{1}{T_H} + \left\{
\begin{array}{ll}
\sqrt{\frac{f}{E}} \,, & \mbox{for open strings} \,, \\
\frac{(\beta_H E)^{(2d - 1)}}{(2d - 1)!} \frac{\ell_s^2}{R^2}
e^{-\ell_s^3 E/R^2}\,,& \mbox{for closed strings}\,,
\end{array} \right. \label{T-E}
 \ee
or invert them to get the energy of the system
\be
E \approx  \left\{ \begin{array}{ll}
\frac{f T^2}{\left(1 - T/T_H\right)^2} \,, & \mbox{for open strings} \,, \\
\frac{R^2}{\ell_s^3}\ln{\left[\frac{\ell_s^3}{R^2}\frac{T}{(1 -
T/T_H)}\right]}\,,& \mbox{for closed strings}\,.\\
\end{array} \right.
 \ee
The specific heat $C_V \equiv \partial E/\partial T$, therefore, is
\be
C^{\rm{closed}}_V \approx \frac{R^2/\ell_s^3}{T \left(1 -
T/T_H\right)}\,, \label{sh_closed}
\ee
for closed strings and
\be  C^{\rm{open}}_V \approx \frac{2 f T}{\left(1 - T/T_H\right)^3}
\,,  \label{sh_open} \ee
for open strings with $f = (R^d/\ell_s^{d - 1})$. In both cases the
specific heat is positive and approach the Hagedorn temperature from
below unlike the nonlimiting cases. For instant, the specific heat
for closed strings in an infinite volume (i.e., non-compact
dimensions)  is negative and approach the Hagedorn temperature from
above~\cite{djnt:1992}
$$
C^{\textrm{nc}}_V = - \frac{d + 2}{2} \left(\frac{T_H}{T -
T_H}\right)^2 \,,
$$
which could be a sign of phase transition or no thermal equilibrium.
For comparison, note that the specific heat of an ideal gas of
relativistic point particles in a large volume $V = R^d$ is always
positive
\be
C^{\textrm{particle}}_V \approx (d + 1) \left(\frac{d}{d +
1}\right)^d R^d T^d \label{sheat_particle} \,.
\ee
Note that the specific heat for open strings and for massless
relativistic point particles scales similarly like $R^d$, while for
closed strings scales as $R^2$ in any number of spatial dimensions
larger than two.  This is a crucial point and later on will play an
important role to get the scale-invariant spectrum in $d = 3$

\section{String Gas Cosmology: Thermal Fluctuations}

Now, let's calculate the fluctuations of the energy-momentum tensor
during the Hagedorn phase.

For a warm up let's recall that the average of a thermodynamical
quantity, ${\cal H}$ can be derived from the partition function $Z$
through
\be \langle {\cal H} \rangle = - \frac{\partial \ln{Z}}{\partial
\beta}\,. \ee
Taking another derivative of $Z$ will give us
\be \langle {\cal H}^2 \rangle = \langle {\cal H} \rangle^2 +
\frac{\partial^2 \ln{Z}}{\partial \beta^2}\,. \ee
The fluctuation in ${\cal H}$ is
\be \delta {\cal H} = {\cal H} - \langle {\cal H} \rangle \,, \ee
and thus the mean square is
\be \langle \delta {\cal H}^2 \rangle = \langle ({\cal H} - \langle
{\cal H} \rangle)^2 \rangle = \langle {\cal H}^2 \rangle - \langle
{\cal H} \rangle^2 = \frac{\partial^2 \ln{Z}}{\partial \beta^2}\,.
\ee
By the same fiat, I can find the average of an operator ${\cal
O^\alpha}$ by
\be \langle {\cal O^\alpha} \rangle  = \frac{\partial
\ln{Z}}{\partial {\cal Q^\alpha}} \,, \ee
where ${\cal Q^\alpha}$ is the conjugate to ${\cal O^\alpha}$.
Taking another derivative with respect to ${\cal Q^\alpha}$  yields
\be \langle {\cal O^\alpha O^\beta} \rangle  = \langle {\cal
O^\alpha\rangle\langle O^\beta} \rangle + \frac{\partial^2
\ln{Z}}{\partial {\cal Q^\alpha Q^\beta}}\,. \ee
The mean square fluctuation is, hence
\be \langle \delta {\cal O^\alpha} \delta {\cal O^\beta} \rangle  =
\langle {\cal O^\alpha O^\beta} \rangle  - \langle {\cal
O^\alpha\rangle\langle O^\beta} \rangle = \frac{\partial^2
\ln{Z}}{\partial {\cal Q^\alpha \partial Q^\beta}}\,, \ee
where $\delta {\cal O^\alpha} = {\cal O^\alpha} - \langle {\cal
O^\alpha} \rangle$ is the fluctuation from the mean.

The mean energy-momentum tensor $\langle T^\mu{}_\nu \rangle$ can be
defined by
\be \langle T^\mu{}_\nu \rangle = 2 \frac{G^{\mu \alpha}}{\sqrt{ -
G}}\frac{\partial \ln{Z}}{\partial G^{\alpha\nu}}\,, \ee
where $G_{\mu\nu}$ is the metric.  Taking another derivative of $Z$
gives,
\begin{widetext}
\be
\langle T^\mu{}_\nu T^\sigma{}_\lambda \rangle  =  \langle
T^\mu{}_\nu \rangle \langle T^\sigma{}_\lambda \rangle +  2
\frac{G^{\mu \alpha}}{\sqrt{ - G}}\frac{\partial}{\partial
G^{\alpha\nu}}\left(\frac{G^{\sigma \delta}}{\sqrt{ -
G}}\frac{\partial \ln{Z}}{\partial G^{\delta\lambda}}\right)+ 2
\frac{G^{\sigma \alpha}}{\sqrt{ - G}}\frac{\partial}{\partial
G^{\alpha\lambda}}\left(\frac{G^{\mu \delta}}{\sqrt{ -
G}}\frac{\partial \ln{Z}}{\partial G^{\delta\nu}}\right)\,,
\ee
\end{widetext}
and thus the mean square fluctuation is
\begin{widetext}
\ba
C^\mu{}_\nu{}^\sigma{}_\lambda  =  \langle \delta T^\mu{}_\nu
\delta T^\sigma{}_\lambda \rangle & = & \langle T^\mu{}_\nu
T^\sigma{}_\lambda \rangle  - \langle T^\mu{}_\nu \rangle \langle
T^\sigma{}_\lambda \rangle  \nonumber \\
& = & 2 \frac{G^{\mu \alpha}}{\sqrt{ - G}}\frac{\partial}{\partial
G^{\alpha\nu}}\left(\frac{G^{\sigma \delta}}{\sqrt{ -
G}}\frac{\partial \ln{Z}}{\partial G^{\delta\lambda}}\right) + 2
\frac{G^{\sigma \alpha}}{\sqrt{ - G}}\frac{\partial}{\partial
G^{\alpha\lambda}}\left(\frac{G^{\mu \delta}}{\sqrt{ -
G}}\frac{\partial \ln{Z}}{\partial G^{\delta\nu}}\right) \,,
\ea
\end{widetext}
with $\delta T^\mu{}_\nu = T^\mu{}_\nu - \langle T^\mu{}_\nu
\rangle$.

\begin{figure}
\includegraphics[height=6cm]{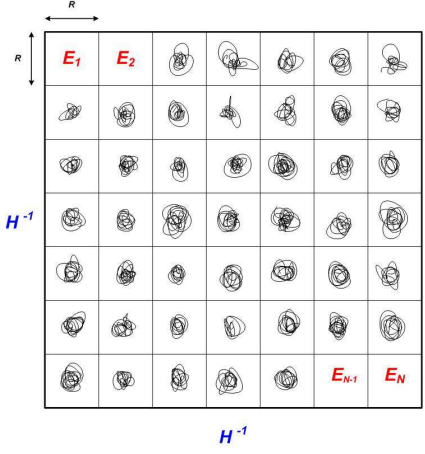}
\caption{Kayhaanistan: the block universe which consists of many
Kayhaanaks (mini-cosmoses) of linear size $R$.  The thermal
correlation function of thermal fluctuations are calculated in each
block.} \label{fig:2}
\end{figure}

In order to make our thermodynamical approach sensible,  we divide
the spacetime inside the Hubble radius ({\it Kayhaanistan}
\footnote{Kayhaanistan is a Persian word and means a place for many
mini cosmoses or {\it Kayhaanaks}}), $H^{-1}$, to small blocks ({\it
Kayhaanaks}) of size $\ell_s \ll R \ll H^{-1}$, where $R$ is almost
independent of time during the Hagedorn phase (see Figure
\ref{fig:2}). Now in each of these Kayhaanaks the energy (mass)
fluctuations can be calculated. Mass fluctuation inside of each
Kayhaanak is responsible for the observed fluctuations in the cosmic
microwave background radiation (CMBR). The partition function $Z =
\exp{(-\beta F)}$, where $F = F(\beta\sqrt{-G_{00}}, R)$ is the
string free energy with $\beta \sqrt{-G_{00}} =
T^{-1}\sqrt{-G_{00}}$ acting as the Euclidean time. Therefore
$C^0{}_0{}^0{}_0$, becomes \footnote{See \cite{Brandenberger:2006xi}
for the calculation of other components of the correlation
function.}
\ba
C^0{}_0{}^0{}_0 & = & \langle \delta\rho^2 \rangle  =  \langle
\rho^2 \rangle - \langle \rho \rangle ^2 \nonumber \\
& = & - \frac{1}{R^{2 d}} \frac{\partial}{\partial \beta}\left(F +
\beta \frac{\partial F}{\partial \beta}\right)\nonumber \\
& \equiv & - \frac{1}{R^{2 d}}\frac{\partial E}{\partial \beta}
\equiv \frac{T^2}{R^{2d}} C_V \,,
\ea
where, as before, $V = R^d$ is the spatial volume and $E =
F+\beta(\partial F/\partial \beta)$ is the total energy.

Now using the result that we obtained in (\ref{sh_closed}) and
(\ref{sh_open}) give us the following results for the mean energy
density fluctuations squared
\be
\langle \delta \rho^2 \rangle_{\rm{closed}} \approx
\frac{R^{-2(d - 1)}}{\ell_s^3}\frac{T}{(1 - T/T_H)}\,,
\ee
for closed strings and
\be
\langle \delta \rho^2 \rangle_{\rm{open}} \approx \frac{2
R^{-d}}{\ell_s^{(d - 1)}} \frac{T^3}{(1 - T/T_H)^3}\,,
\ee
for open strings.  As far as the length scaling is concerned,
$\langle \delta \rho^2 \rangle$ for open strings is same as the one
for relativistic massless point particles according to
(\ref{sheat_particle})
\be
\langle \delta \rho^2 \rangle_{\rm{particle}} \sim R^{-d} T^d
\,.
\ee

\section{The Power Spectrum of the Generated Perturbations}
In order to compute the power spectrum of these fluctuations, we
will use the theory of linear cosmological perturbations around a
four-dimensional homogeneous and isotropic cosmology
\cite{MFB,RHBrev4,BST,BKP,RHB}.

The key point here is the fact that the thermal fluctuations
generated during the Hagedorn phase are well inside the Hubble
radius and exit the Hubble radius at a time very close to the
transition time from Hagedorn era to the radiation phase.  These
fluctuations then will reenter the Hubble radius at some later time
well within radiation dominated epoch.

In the absence of anisotropic stress, there is only one physical
degree of freedom, namely the relativistic generalization of the
Newtonian gravitational potential. In a flat universe in the
conformal Newtonian gauge, the metric takes the form
\be
ds^2 \, = \, - (1 + 2 \Phi) dt^2 + a(t)^2 (1 - 2 \Phi) d{\bf
x}^2 \, ,\label{pert_metric}
\ee
where $t$ is physical time, ${\bf x}$ are the comoving spatial
coordinates of the three large spatial dimensions, $a(t)$ is the
cosmological scale factor and $\Phi({\bf x}, t) \ll 1$ represents
the scalar fluctuation mode in the gravitational potential.  Here we
discount the tensor modes (see \cite{Brandenberger:2006xi} for the
discussion of the tensor modes in the Hagedorn phase).

For the metric (\ref{pert_metric}) with small perturbations, the
Einstein and energy-momentum tensor can be split to unperturbed and
linear terms in fluctuations.  The linearized equations for the
perturbations are
\be
\delta G^\alpha{}_\beta = \kappa_{d+1}^2 \delta T^\alpha{}_\beta
\,,
\ee
where $\kappa_{d+1}^2$ is the Einstein gravitational coupling
constant in an arbitrary spatial dimension $d$
\cite{Mansouri:1996ik}
$$
\kappa^2_{d+1} = \frac{2(d -1) \pi^{d/2} G_{d +1}}{(d -
2)\Gamma(d/2)}\,.
$$

\begin{figure}
\includegraphics[height=10cm]{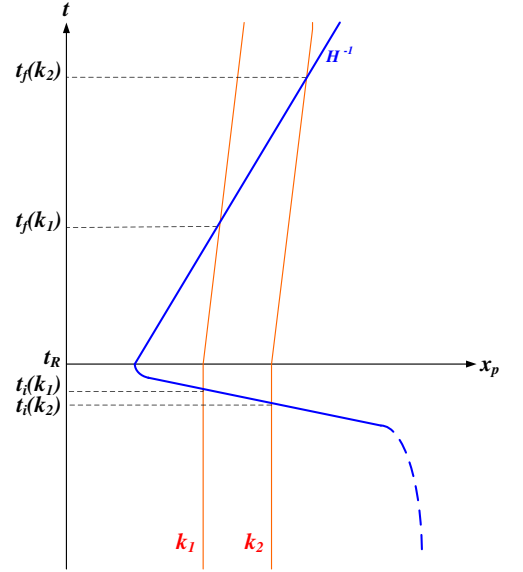}
\caption{Space-time diagram (sketch) showing the evolution of fixed
comoving scales. The vertical axis is time, the horizontal axis is
physical distance. The Hagedorn phase ends at the time $t_R$ and is
followed by the radiation-dominated phase of standard cosmology. The
solid curve represents the Hubble radius $H^{-1}$ which is
cosmological and almost constant during the quasi-static Hagedorn
phase, shrinks abruptly to a microphysical scale at $t_R$ and then
increases linearly in time for $t > t_R$. Fixed comoving scales
(labeled by $k_1$ and $k_2$) which are currently probed in
cosmological observations have wavelengths which are smaller than
the Hubble radius during the Hagedorn phase. They exit the radius at
times $t_i(k)$ just prior to $t_R$, and propagate with a wavelength
larger than the Hubble radius until they reenter the Hubble radius
at times $t_f(k)$.} \label{fig:3}
\end{figure}

The short-wavelength perturbations are very important as they are
the reason for the acoustic peaks in the cosmic microwave background
(CMB) spectrum. On scales smaller than the Hubble radius, the
gravitational potential $\Phi$ is determined by the matter
fluctuations via the Einstein constraint equation ($\delta G^0{}_0 =
\kappa_{d+1}^2 \delta T^0{}_0$) which is the relativistic
generalization of the Poisson equation of Newtonian gravitational
perturbation theory:
\be
\nabla^2 \Phi =  \frac{2 \pi^{d/2} G_{d+1}} {\Gamma[d/2]} \delta
\rho \,, \label{poisson_real}
\ee
where $G_{d+1}$ is the Newton's Gravitational coupling constant in
$d$ compact but yet large spatial dimensions and $\Gamma[x] \equiv
(x -1)!$.  Using Fourier transform of (\ref{poisson_real}) we find
that at the Hubble radius crossing time
\be
|\Phi_k|^2 =  \frac{4 \pi^d G^2_{d+1}} {\Gamma^2[d/2]} k^{-4}
\langle \delta \rho^2_k \rangle \label{poisson_k}\,.
\ee
By (\ref{poisson_k}) one can calculate the power spectrum of the
metric fluctuations $\Phi_k$ at the end of the Hagedorn phase or to
be more precise, at the time $t_i(k)$ when the fluctuate mode
labeled by $k$ exits the Hubble radius (see Figure \ref{fig:3}). In
the context of standard cosmological perturbation theory, for modes
with $k/a(t) \ll H(t)$, one can find a conserved quantity
$\mathcal{R}$ which is related to the spatial curvature in comoving
slicing of spacetime and in Newtonian gauge (in the absence of
anisotropic pressure) is given by \cite{Weinberg:2003sw}
\be
\mathcal{R}= \zeta + \left(\frac{k^2}{3 a^2 \dot
H}\right)\Phi\,,
\ee
where
\be
\zeta = - \Phi + \frac{\delta \rho}{3(\rho + p)}= - \Phi +
\frac{\delta \rho}{3 \rho (1 + w)} \,
\ee
is the spatial curvature on the constant energy density spacelike
surface and $w = p/\rho$ is the equation of state.  Note that for
the super-Hubble radius fluctuations, i.e., in the limit of $k
\rightarrow 0$, $\mathcal{R} \rightarrow \zeta$.  The conservation
of $\mathcal{R}$ (or $\zeta$) means conservation of $\Phi$ on
super-Hubble scale as far as the equation of state of the background
does not change drastically from one phase to another one. For
string gas cosmology the equation of state at the end of the
Hagedorn phase changes from $w \approx 0$ to $w \approx 1/3$ at the
beginning of the radiation dominated era which is much milder than a
change compare to the cosmic inflationary scenario.  As long as the
equations of four space-time dimensional general relativistic
cosmological perturbation theory apply, then $\Phi$ is conserved on
super-Hubble scales as long as the equation of state of the
background does not change significantly. The use of ordinary
general relativistic cosmological perturbation theory is certainly
justified after the end of the Hagedorn phase, but not necessarily
in the time interval between mode exiting from the Hagedorn phase
and radiation dominated epoch.

Assuming the validity of the arguments of the previous paragraphs by
ignoring the contribution of the dilation fluctuation, then the
scalar spectral index $n_s$ of the cosmological perturbations can be
determined.

The dimensionless power spectrum, in $d$ spatial dimensions is given
by \cite{Peacock:1999}
\be
\Delta^2(k) = \frac{V}{(2\pi)^d} \frac{2 \pi^{d/2}}
{\Gamma[d/2]} k^d P(k)  \,,
\ee
where $P_{\delta\rho_k}= \langle \delta\rho_k^2\rangle$ and
$P_{\Phi_k} = |\Phi_k|^2$ are the power spectra of the energy
density fluctuations and the metric perturbation, respectively. Thus
the dimensionless power spectrum of the metric fluctuation is
\be
\Delta^2_{\Phi_k}(k)  =   \frac{4 \pi^d G^2_{d+1}}
{\Gamma^2[d/2]} k^{-4} \Delta^2_{\delta\rho_k}(k)\,,
\ee
which has the following forms at the end of Hagedorn regime
\be
\Delta^2_{{\Phi_k}_{cl}}(k) \simeq  \frac{\pi^{(d/2)} G^2_{d+1}}
{2^{(d - 3)}\Gamma^3[d/2]\ell_s^3}\frac{T}{(1 - T/T_H)} k^{2(d - 3)}
\label{closed_power} \,,
\ee
for closed strings and
\be
\Delta^2_{{\Phi_k}_{op}}(k) \simeq  \frac{\pi^{(d/2)} G^2_{d+1}}
{2^{(d - 4)}\Gamma^3[d/2]\ell_s^{d - 1}} \frac{T^3}{(1 - T/T_H)^3}
k^{d - 4} \label{open_power} \,,
\ee
for open strings.

Obviously, as one can see from (\ref{closed_power}) and
(\ref{open_power}), the power spectrum is independent of the length
of fluctuations  in $d = 3$ and $d = 4$ for closed and open strings,
respectively. Therefore, whereas only closed strings provide us a
scale-invariant spectrum in $D = 3 + 1$ dimensions,
\be
\Delta^2_{{\Phi_k}_{cl}}(k) \simeq \frac{8 G_4^2}{\ell_s^3}
\frac{T}{(1 - T/T_H)} \label{scale_inv}\,,
\ee
the spectrum of open strings in $d = 3$ slopes towards red (more
power on large scales)
\be
\Delta^2_{{\Phi_k}_{cl}}(k) \simeq \frac{4 G_4^2}{\ell_s^2}
\frac{T^3}{k (1 - T/T_H)^3} \label{scale_red}\,.
\ee
Note that (\ref{scale_inv}) near the Hagedorn temperature $T_H (=
1/4\pi\ell_s) - T \approx \mathcal{O}(<10^{-2})$ corresponds to a
hierarchy of $(\ell_{Pl}/\ell_s) \sim 10^{-4} - 10^{-3}$ for an
observed amplitude of $10^{-5}$.  On the other hand, the ratio
$(\ell_{Pl}/\ell_s)$ is proportional to the string coupling
constant, $g_s$. Our assumption from the beginning was that $g_s \ll
1$.  It seems that our result is consistent with that assumption.

Finally, notice that the spectrum (\ref{scale_inv}) looks
scale-invariant to the first approximation (assuming $T$ is
independent of $k$), but since $T(t_{exit}(k))$ at time
$t_{exit}(k)$ depends on $k$, the power spectrum gains a slight {\it
red tilt} due to the $1 - T(t_{exit}(k))/T_H$ in the denominator.
Since the exact form of $k$ dependency of $T$ is hard to find due to
our lack of knowledge about phase transition at the end of Hagedorn
epoch, one possible ansatz for the factor in the denominator,
noticing that $T(t_{exit}(k))$ is a slowly decreasing function of
$k$ (shorter modes or larger $k$ modes exit Hubble radius later), is
\be
1 - \frac{T(t_{exit}(k))}{T_H} \approx \alpha
\left(\frac{k}{k_0}\right)^\epsilon \,,
\ee
where both $\alpha$ and $\epsilon$ are positive numbers much smaller
than unity for some constant $k_0$.  Therefore the scaler spectral
index $n_s$ is
\be
n_s - 1 \approx - \epsilon
\,,\ee
which yields $n_s \approx 0.95$ if we assume $\epsilon \approx 0.05$
to be compatible with the third year WMAP result
\cite{Spergel:2006hy}. So the spectrum has a mild red tilt depending
on the value of $\epsilon$.

\section{Conclusion}
In this paper, we have studied the generation and evolution of
cosmological fluctuations in a model of string gas cosmology in
which an early quasi-static Hagedorn phase is followed by the
radiation-dominated phase of standard cosmology, without an
intervening period of inflation. Due to the fact that the Hubble
radius during the Hagedorn phase is cosmological, it is possible to
produce fluctuations using causal physics. Assuming thermal
equilibrium on scales smaller than the Hubble radius, we have used
string thermodynamics to study the amplitude of density fluctuations
during the Hagedorn phase. The mean square energy fluctuations are
determined by the specific heat of the string gas. To compute the
perturbations on a physical length scale $R$, we apply string
thermodynamics to a box of size $R$. Working under the assumption
that all spatial dimensions are compact (but sufficiently large),
the specific heat turns out to scale as $R^2$ for closed strings.
This is an intrinsically stringy effect: in the case of point
particle thermodynamics, the specific heat would scale as $R^d$. The
$R^2$ scaling of the specific heat leads to a scale-invariant
spectrum of metric fluctuations.  The compactness of the spatial
dimensions are very crucial here.  If we were dealing with
noncompact dimensions, instead, the specific heat would have been
negative. The positiveness of the specific heat for closed strings
in compact spatial dimensions, among other things, can assure us the
absence of formation of any primordial black holes in the Hagedorn
phase, for instance.

Although our cosmological scenario provides a new mechanism for
generating a scale-invariant spectrum of cosmological perturbations,
it does not solve all of the problems which inflation solves. In
particular, it does not solve the flatness problem. Without assuming
that the there are some large spatial dimensions are much larger
than the string scale, we do not obtain a universe which is
sufficiently large today.  The longevity of the Hagedorn phase,
however, could help us to explain the homogeneity and the absence of
$U(1)$ monopoles.  During the Hagedorn phase, closed strings would
have enough time to travel the entire universe and communicate with
strings of other parts of the universe so that the whole universe
becomes almost homogenous.  On the other hand, if monopoles form
during the Hagedorn phase, perhaps there would be enough time for
them to annihilate each other.  These issues, of course, need more
detailed studies and will be discussed in future publications.

Our scenario may well be testable observationally. Taking into
account the fact that the temperature $T$ evaluated at the time
$t_i(k)$ when the scale $k$ exits the Hubble radius depends slightly
on $k$, the formula (\ref{scale_inv}) leads to a calculable
deviation of the spectrum from exact scale-invariance. Since
$T(t_i(k))$ is decreasing as $k$ increases, a slightly red spectrum
is predicted. Since the equation of state does not change by orders
of magnitude during the transition between the initial phase and the
radiation-dominated phase as it does in inflationary cosmology, the
spectrum of tensor modes is not expected to be suppressed compared
to that of scalar modes.

Finally, it would interesting to study the fluctuation of the
dilatonic field and its contribution to the metric fluctuation
during the Hagedorn phase in more details.  We hope we can address
this and other related issued in future publications.

\section*{Acknowledgements}
I really am thankful to R. Brandenberger, C. Vafa and L. Motl
without whom this work could no be accomplished. I also would like
thank R. Allahverdi, A. Guth, J. Khury, L. Kofman, A. Linde, V.~F.
Mukhanov, S. Patil, A. Peat, P. Ramond, C. Thorn and S. Watson for
many useful discussions. The work of A.N. is supported in part by
NSF grant PHY-0244821 and DMS-0244464


\end{document}